\documentclass[preprint2]{aastex}
\usepackage{amsmath}
\usepackage{amssymb}
\usepackage{graphicx}

\usepackage{natbib}
\def\aap{A\&A} 
\def\apj{ApJ} 
\def\apjl{ApJL} 
\def\apjs{ApJS} 
\def\mnras{MNRAS} 
\def\nat{Nature} 
\def\prd{Phys.~Rev.~D} 

\def\eg{{e.g.}} 

\def\Ms{{\rm M}\s} 
\def\yr{{\rm yr}} 
\def\GHz{{\rm GHz}} 
\def\m{{\rm m}} 
\def\mum{\mu\m} 
\def\mm{{\rm m}\m} 
\def\pc{{\rm pc}} 
\def\kpc{{\rm k}\pc} 
\def\Ms{M_\odot} 
\def\muas{\mu{\rm as}} 

\def\e{{\rm e}}
\def\SgrA{{Sgr A*}}

\title{On The Nature of the Compact Dark Mass at the Galactic Center}

\author{Avery E. Broderick\altaffilmark{1} and Ramesh Narayan\altaffilmark{2}}
\affil{Institute for Theory and Computation, Harvard-Smithsonian Center
for Astrophysics, MS 51, 60 Garden Street, Cambridge, MA 02138, USA}
\altaffiltext{1}{abroderick@cfa.harvard.edu}
\altaffiltext{2}{rnarayan@cfa.harvard.edu}

\shorttitle{The Dark Mass in the Galactic Center}
\shortauthors{A.~E. Broderick \& R. Narayan}

\begin{document}

\begin{abstract}

We consider a model in which \SgrA, the $3.7\times10^6\,\Ms$
supermassive black hole candidate at the Galactic Center, is a compact
object with a thermally emitting surface.  For very compact surfaces
within the photon orbit, the thermal assumption is likely to be a good
approximation because of the large number of rays which are strongly
gravitationally lensed back onto the surface.  Given the very low
quiescent luminosity of {\SgrA} in the near infrared, the existence of
a hard surface, even in the limit in which the radius approaches the
horizon, places a severe constraint upon the steady mass accretion
rate onto the source: $\dot{M}\lesssim10^{-12}\,\Ms\yr^{-1}$.  This
limit is well below the minimum accretion rate needed to power the
observed submillimeter luminosity of \SgrA: $\dot{M} >
10^{-10}\,\Ms\yr^{-1}$.  We thus argue that \SgrA~does not have a
surface, i.e., it must have an event horizon.  The argument could be
made more restrictive by an order of magnitude with $\muas$ resolution
imaging, \eg, with submillimeter very-long baseline interferometry.
\end{abstract}


\maketitle

\section{Introduction}

Infrared observations of individual stars in the Galactic Center imply
the existence of a dark object of mass $M \approx 3.7\times10^6\,\Ms$,
constrained to lie within $45\,{\rm AU}$ (or $10^3 GM/c^2$) of the
radio source \SgrA~\citep{Scho_et_al:03,Ghez_et_al:05,Eise_et_al:05}.
Observations at $3.5\,\mm$ and $7\,\mm$ further constrain the extent
of the radio emission from \SgrA~to less than $1-2\,{\rm AU}$ (or
$10-20GM/c^2$) \citep{Shen_etal:05,Bowe_etal:04}.  The favored
interpretation of these observations is that \SgrA~is a supermassive
black hole.  Indeed, the current constraints rule out many alternative
explanations, including clusters of stellar mass compact objects
\citep{Maoz:98} and fermion balls \citep{Scho_etal:02}.  Nonetheless, it remains to be conclusively
demonstrated that the dark mass at the Galactic Center is a true black
hole with an event horizon.

If {\SgrA} is not a black hole, then it must have a surface at some
radius $R$.  Although general relativistic considerations coupled with
reasonable assumptions on the equation of state of matter require $R
\geq 9GM/4c^2$ \citep[see, \eg,][]{Shap-Teuk:86}, alternatives to
general relativity exist which allow smaller radii, despite the fact
that the exterior spacetime may be arbitrarily close to that predicted
by general relativity (\eg, scalar-tensor theories,
\citealt{Fuji-Maed:03}, gravastars, \citealt{Mazu-Mott:01}, boson stars
\citealt{Torr-Capo-Lamb:00}).
Thus, in principle, $R$ could have any value greater than $2GM/c^2$, the horizon
radius (we restrict our analysis to non-spinning objects).  In this
{\it Letter}, we show that current observations do not favor {\SgrA}
having such a surface.

We assume that any putative surface of {\SgrA} is in steady state in
the presence of accreting gas, and that it emits the accreted energy
thermally.  The latter assumption is reasonable since, even for models
in which the radius of the surface approaches $2GM/c^2$, the
thermalization timescale is short in comparison to the lifetime of
{\SgrA} (or of an observer)\footnote{The thermalization timescale is
expected to be on the order of the cooling time of the infalling
matter, which is not very different from the free-fall time-scale
$\sim 100$ s at the surface.  Since the timescale diverges only
logarithmically as measured at infinity even for extremely compact
configurations (\eg, the gravastar), the thermalization time is
increased by only a few orders of magnitude.}.  Indeed, for surfaces
contained well within the photon orbit, strong gravitational lensing
significantly decreases the number of photon trajectories that escape
to infinity.  Most outgoing rays return to be absorbed by other parts
of the surface, so that the object will resemble the classical
example of a blackbody: a thermal cavity with a pinhole.

If \SgrA~accretes at the Bondi rate from the hot gas surrounding it,
the accretion rate is expected to be $\dot M \sim 10^{-6}
~M_\odot\,{\rm yr^{-1}}$ \citep{Baga_etal:03}.  A more likely
scenario is that the source accretes via a radiatively inefficient
accretion flow \citep[RIAF;][and references
therein]{Nara-Yi-Maha:95,Yuan-Quat-Nara:03}, with a mass accretion
rate in the range $\dot M \sim 10^{-8.5}-10^{-6} ~M_\odot\,{\rm
yr^{-1}}$.  In a RIAF model, essentially all the potential energy
released by the accreting gas would be radiated from the surface of
{\SgrA} (assuming it has a surface), with a predicted luminosity at
infinity of
\begin{equation}
L_{\rm surf} \approx \eta \dot M c^2 ,\label{Lsurf}
\end{equation}
where the efficiency factor $\eta$ is the fraction of the rest mass
energy of the infalling gas that is released as radiation.  In the
Newtonian case, $\eta$ is simply $GM/c^2R$.  In general relativity,
$\eta$ is given in terms of the gravitational redshift $z$ at the
surface:
\begin{equation}
\eta = z/(1+z), \quad 1+z = (1-2GM/c^2R)^{-1/2}. \label{eta}
\end{equation}
Although it is highly unlikely that \SgrA~has a radiatively {\it
efficient} accretion disk, even such a model requires a fairly large
$\dot M$.  For example, the observed bolometric luminosity of {\SgrA}
of $10^{36} ~{\rm erg\,s^{-1}}$ implies a minimum accretion rate of
$\dot M \sim 2\times 10^{-10} ~M_\odot\,{\rm yr^{-1}}$ for a radiative
efficiency of 10\%.  To within a factor of a few (depending on the
nature of the boundary layer at the inner edge of the disk), the
luminosity from the surface of {\SgrA} is again predicted to be $\sim
\eta\dot Mc^2$ with $\eta$ not very different from (\ref{eta}).  All
the above estimates are for gas accretion.  When stellar capture
events are considered, the average accretion rate can be as high as
$10^{-5}~M_\odot\,\yr^{-1}$ to $10^{-3}~M_\odot\,\yr^{-1}$, though
this would be expected to be in the form of transient accretion events
\citep{Mago-Trem:99}.  Note that the $\dot M$ estimates given here are
from the point of view of a distant observer, i.e., they represent the
rate of accretion of rest mass per unit time as measured at infinity.

In \S~2, we derive upper limits on $\dot M$ from the observed
near-infrared (NIR) fluxes of {\SgrA} and compare these with the
various estimates of $\dot M$ given above.  On this basis we argue
that \SgrA~is unlikely to have a surface and therefore that it must be
a black hole.  In \S~3, we present theoretical images of the RIAF
model discussed in \citet{Yuan-Quat-Nara:03} and
\citet{Brod-Loeb:05b}, and show that imaging experiments alone cannot
distinguish between a black hole and a compact object with a surface.
However, by combining imaging with the argument presented in \S~2, we
show that one could make the case for an event horizon stronger.  We
conclude in \S~4 with a discussion.  In what follows, unless otherwise
noted, we use geometrized units ($G=c=1$).

\section{NIR Limits on the Mass Accretion Rate}

The observed flux $F_\nu$ at a frequency $\nu$ from a surface emitting
thermal blackbody radiation is simply the blackbody spectrum
multiplied by the apparent solid angle of the surface on the sky.  For
a thermally emitting compact spherical surface at the Galactic Center
this is
\begin{equation}
F_\nu = \frac{2 h \nu^3}{c^2} \frac{\e^{-h\nu/kT}}{1-e^{-h\nu/kT}}
\frac{\pi b^2}{D^2}\,, \label{Fnu}
\end{equation}
where $T$ is the blackbody temperature as measured by the observer (at
infinity), $D\simeq8\,\kpc$ is the distance to the Galactic Center,
and the apparent size $b$ of the radiating surface is given in terms
of its radius $R$ by
\begin{equation}
b^2 =
\left\{
\begin{aligned}
27 M^2, && R<3M ,\\
R^3/(R-2M), && R\ge3M .
\end{aligned}
\right.\label{b2}
\end{equation}
Equation (4) includes the general relativistic correction due to
strong lensing.  Note that, for $R$ inside the photon orbit $3M$, the
apparent size $b$ no longer decreases with decreasing $R$.

For a given choice of radius $R$ and a given upper limit on the flux
$F_\nu$, equations (\ref{Fnu}) and (\ref{b2}) provide an upper limit
$T_{\rm max}$ on the observed temperature (at infinity) and hence a
limit on the surface luminosity as measured at infinity: $L_{\rm surf}
< 4\pi b^2 \sigma T_{\rm max}^4$.  Then, from equations (\ref{Lsurf})
and (\ref{eta}) we obtain the following upper limit on the mass
accretion rate on the surface,
\begin{equation}
\dot{M} < 4\pi b^2\sigma T_{\rm max}^4/c^2 \eta \,. \label{Mdot}
\end{equation}
Thus, for each flux measurement $F_\nu$ of {\SgrA} and an assumed
radius $R$, we obtain an estimate of $\dot M_{\rm max}(R)$, the
maximum accretion rate for that $R$.  This limit, which assumes
nothing more than blackbody emission, can be compared with the
mass accretion rates which are required to explain the observed
luminosity and spectra of {\SgrA} (discussed in \S~1).

\begin{deluxetable}{ccc}
\tablewidth{0pt}
\tablecaption{\label{flux_limits}
Near-Infrared Flux Limits of \SgrA}
\tablehead{
\colhead{$\lambda\,(\mum)$} &
\colhead{$F_\nu\,({\rm mJy})$} &
\colhead{Ref.}}
\startdata
1.6 & $11$ & \citet{Stol_etal:03}\\
2.1 & $2.8$ & \citet{Ghez_etal:05b}\\
3.8 & $1.28$ & \citet{Ghez_etal:05b}\\
4.8 & $3.5$ & \citet{Clen_etal:04}\\
\enddata
\end{deluxetable}

\begin{figure}[t!]
\begin{center}
\includegraphics[width=\columnwidth]{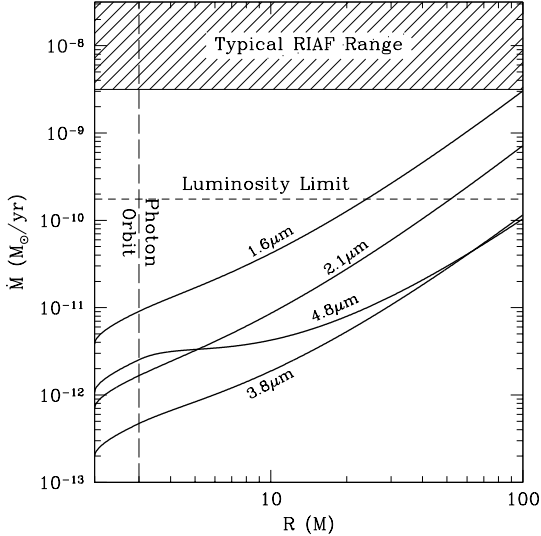}
\end{center}
\caption{The four curves show upper limits on the mass accretion rate
of {\SgrA} as a function of the surface radius $R$, derived from the
observed quiescent fluxes at $1.6$, $2.1$, $3.8$ and $4.8\,\mum$
listed in Table \ref{flux_limits}.  The surface is assumed to radiate
as a blackbody.  For reference, the photon orbit $R=3M$ is shown by
the vertical dashed line.  The cross-hatched area at the top
corresponds to typical mass accretion rates in RIAF models of \SgrA,
and the horizontal dashed line represents the minimum accretion rate
needed to power the bolometric luminosity of \SgrA (see \S~1).}
\label{mdot_limits}
\end{figure}

In practice, the strongest limits are placed by observations in the
NIR, since the postulated thermal emission from the surface of {\SgrA}
peaks in this region of the spectrum.  The most constraining
observations are listed in Table \ref{flux_limits}, and the
corresponding limits on $\dot M(R)$ are plotted in Figure
\ref{mdot_limits}.  Note that the thermal emission from the surface
peaks in the NIR for $R\sim10M$, whereas for larger radii the surface
is sufficiently cool that the observed fluxes fall in the Wien regime,
resulting in a weakening of the limit on $T_{\rm max}$ and hence on
$\dot M$.  The difference in character between the limit imposed by
observations at $4.8\,\mum$ and the other wavelengths is due to the
fact that the former passes into the Rayleigh-Jeans regime at a larger
radius.  The overall limit on $\dot M$ is simply given by the lower
envelope of all four curves in Figure \ref{mdot_limits}; the envelope
is dominated by the $3.8\,\mum$ curve.

Recent observations at $3.5\,\mm$ and $7\,\mm$ have claimed to resolve
{\SgrA}, limiting the radius to $R\lesssim10-20\,M$ at these
wavelengths \citep{Shen_etal:05,Bowe_etal:04}.  As seen in Figure
\ref{mdot_limits}, this limit on the radius, coupled with the NIR flux
limits, already restricts the allowed accretion rate in \SgrA~to
$\dot{M}\lesssim10^{-12}\,\Ms/\yr$, if the object has a surface.
Since the derived limit is two orders of magnitude lower than the
lowest $\dot M$ allowed by the observed luminosity of \SgrA ($\sim 2
\times 10^{-10} ~M_\odot\,{\rm yr^{-1}}$), and more than three orders
of magnitude lower than the lowest accretion rate required by RIAF
models (see \S~1), the case for \SgrA~not having a surface, i.e., for
it being a black hole, is very strong.  As is apparent from Figure
\ref{mdot_limits}, observations which can further limit $R$ will place
even stronger constraints upon $\dot{M}$, and further strengthen the
case for a black hole.

\section{Images}
\begin{figure*}[!ht]
\begin{center}
\includegraphics[width=\textwidth]{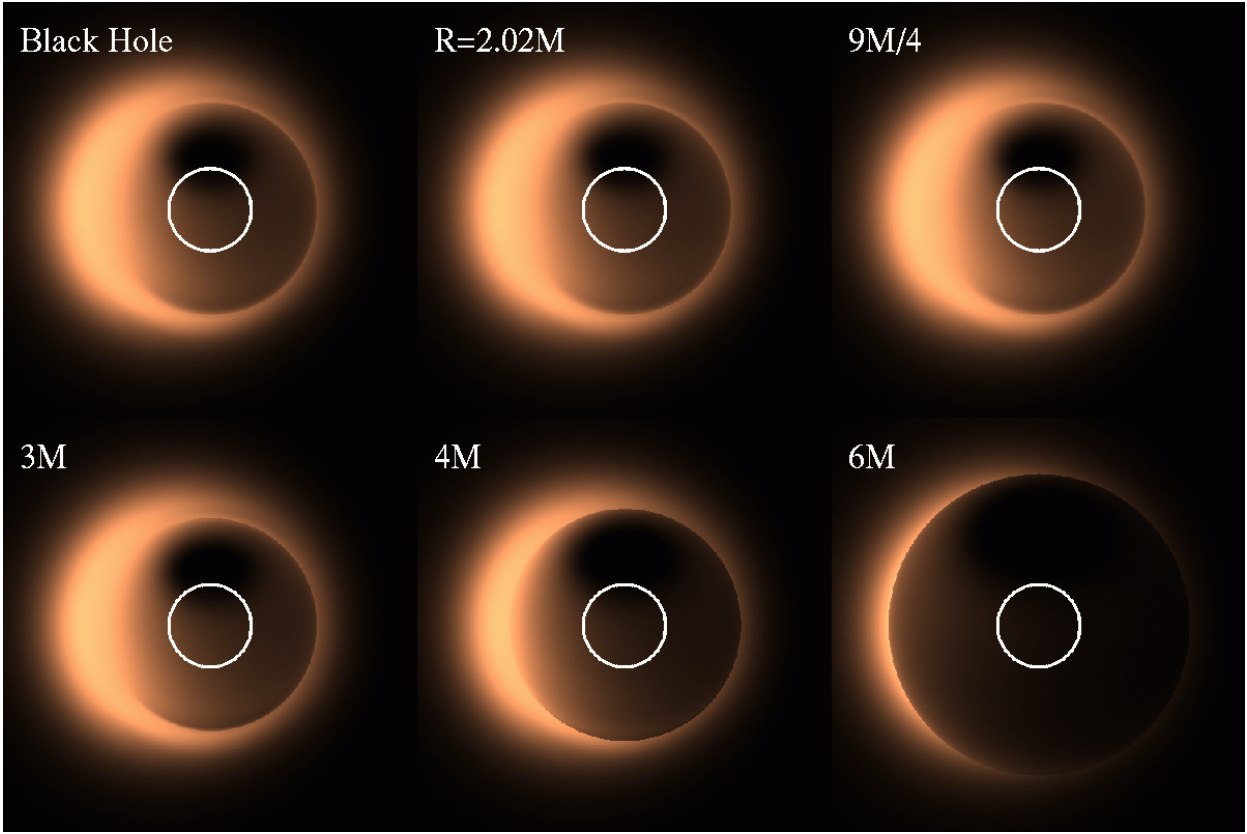}
\end{center}
\caption{$350\,\GHz$ images of a RIAF model of {\SgrA} (the $a=0$
model in \citealt{Brod-Loeb:05b}) in which the gas accretes onto a
compact thermally emitting surface in a Schwarzschild spacetime.  The
panels correspond to different assumed radii of {\SgrA}, given in the
upper left corner of each image, ranging from $1\%$ larger than the
horizon to $6M$.  For comparison, an image of the same accretion flow
onto a black hole is shown in the upper-left panel.  All the images
correspond to a viewing angle $45^\circ$ above the orbital plane.  The
brightness scale is normalized separately in each image, ranging from
maximum intensity to zero (black).  The white circles show the size of
the Schwarzschild horizon $2M$.  Note that the thermally emitting
surface is practically invisible.  This is because of its very low
temperature ($\sim10^4$ K) compared to the brightness temperature of
the relativistic accreting gas ($>10^{10}$ K).}
\label{images}
\end{figure*}

Submillimeter very-long baseline interferometry (VLBI) promises to
provide an angular resolution on the order of $\sim20\,\muas$,
corresponding to a physical scale $\sim5M$ at the Galactic Center.
A significant motivation for developing such a capability is the
prospect of imaging the silhouette or ``shadow'' of the central black
hole
\citep{Falc-Meli-Agol:00,Miyo_etal:04,Doel-Bowe:04,Brod-Loeb:05b,Brod-Loeb:05,Brod-Loeb:05c}.
The shadow is due to strong lensing, and its detection would be a
strong confirmation of one of the major predictions of general
relativity in the limit of strong gravity.  Such imaging will also
constrain the radius of {\SgrA}'s surface and thus strengthen the
argument presented in \S~2.

Figure \ref{images} shows theoretical submillimeter VLBI images of a
RIAF model of {\SgrA} for a number of assumed radii ranging from $1\%$
larger than the horizon to $6M$.  In all cases the mass accretion rate
is taken to be $10^{-8}\,\Ms/\yr$, the canonical value for RIAF models
and orders of magnitude above the limits placed in the preceding
section.  Surface radii in excess of the photon orbit ($3M$) appear as
enlarged silhouettes, as may be seen in the bottom panels.  Thus, high
resolution imaging can immediately limit the size of any radiating
surface in {\SgrA} to less than $3M$.  Combined with the NIR flux
limits, this would strengthen the limit upon the mass accretion rate
by nearly another order of magnitude, making a conclusive case for an
event horizon.

Interestingly, we see that for radii inside of the photon orbit, the
images are nearly indistinguishable from the case in which a horizon
is present.  This is a direct result of the relatively low temperature
of the radiating surface ($\sim 10^4\,{\rm K}$ as measured at
infinity) coupled with the fact that at submillimeter wavelengths the
emission occurs deep in the Rayleigh-Jeans portion of the spectrum.
Thus, while submillimeter imaging can limit the radius of \SgrA, it
cannot by itself prove that the object is a black hole.  It is only
when the imaging results are combined with the NIR flux measurements
in the manner discussed in \S~2 that a strong case can be made for the
existence of a horizon.

\section{Discussion}

We have shown in this {\it Letter} that current NIR flux limits on
{\SgrA} already place a stringent upper limit upon the mass accretion
rate of this compact object, assuming that the object possesses a
surface.  Comparison of these rates to the observed submillimeter
luminosity of the source, and the implied lower limit on the mass
accretion rate, leads to a serious contradiction, thus providing
strong evidence for the absence of a surface.  This argument for
{\SgrA} being a black hole is robust \citep[see][ and references
therein, for other
evidence]{Nara_Gar_McC97,Nara_Gar_McC02,Nara_Heyl02}.  The argument
applies even when the surface is extremely compact, \eg, as expected
in gravastar models \citep[see, \eg,][]{Mazu-Mott:01}.  Note from
equations (\ref{Lsurf}) and (\ref{eta}) that the extremely large
gravitational redshift at the surface of an ultra-compact object
($R\to 2M$) does not cause a reduction in the luminosity observed at
infinity.  On the contrary, for a given $\dot M$, the observed
luminosity is maximum when the redshift goes to infinity \citep[this
is in contrast to][in which the intrinsic luminosity measured on the
surface, as opposed to the mass accretion rate measured at infinity,
was assumed to be fixed]{Abra-Kluz-Laso:02}.  In addition, the
blackbody assumption on which our argument is based is particularly
well motivated when the redshift is large (\S~1).

Three critical assumptions underly our conclusions: ({\em i}) the
surface is in steady state with respect to the accreting material,
({\em ii}) the surface radiates thermally, and ({\em iii}) general
relativity is an appropriate description of gravity external to the
surface.

As mentioned briefly in \S~1, the assumption of steady state is likely
to be a good one, even for surfaces very near the horizon (including
those which are separated from the horizon by a Planck length, the
minimum scale for which a horizon will not develop).  However, it
should be noted that for a black hole the unradiated binding energy of
the accreting matter contributes to an increase of the black hole's
mass.  Thus a black hole is an explicit example of an accreting
compact object which is not in steady state.

Of more concern is the assumption that the surface emits thermally.
For models in which large-scale correlations play a significant role
(\eg, the gravastar) it is unclear what happens to accreting material.
For instance, it is conceivable that one may obtain coherent emission
with wavelengths comparable to the correlation length of the surface,
which in principle could introduce large deviations from the Planck
spectrum.  While we cannot rule out such a model, we would like to
emphasize that, in general, for the gravastar (or similar) model to
remain a viable alternative to the black hole model of {\SgrA}
necessarily requires an extremely exotic emission mechanism that
deviates enormously from blackbody emission.

Finally, some assumption regarding the description of gravity external
to the surface is necessary to compute the flux due to a compact
surface near a strongly gravitating object.  In the absence of a well
tested alternative, general relativity is the natural choice and this
is what we have selected for our calculations.  While we have
explicitly considered a non-rotating black hole, we expect rotation
(the most obvious extension that one would wish to consider) to
primarily broaden the thermal spectrum, changing our results by no
more than factors of order unity.  Multi-wavelength high-resolution
imaging of flares in the Galactic Center has been proposed as a method
by which the nature of the spacetime surrounding the Galactic Center's
black hole may be probed \citep{Brod-Loeb:05,Brod-Loeb:05c}.

To summarize, in the absence of unknown exotic phenomena, the current
NIR flux measurements already conclusively imply the existence of an
event horizon in the black hole candidate {\SgrA} at the Galactic
Center.

\acknowledgments We would like to thank Avi Loeb for pointing out the
highly enhanced accretion rate of {\SgrA} due to stellar capture
events.  This work was supported in part by NSF grant AST 0307433.
A.E.B. gratefully acknowledges the support of an ITC Fellowship from
Harvard College Observatory.

\end{document}